\documentclass[a4paper,11pt]{article}
\usepackage{jheppub}
\usepackage{amsmath,amssymb}
\usepackage{booktabs}
\usepackage{tabularx}
\usepackage{float}

\newcommand{\dd}{\mathrm d}
\newcommand{\ii}{\mathrm i}
\newcommand{\one}{\mathbf 1}
\newcommand{\Tr}{\operatorname{Tr}}
\newcommand{\Ran}{\operatorname{Ran}}
\newcommand{\cE}{\mathcal E}
\newcommand{\cF}{\mathcal F}
\newcommand{\cA}{\mathcal A}
\newcommand{\cB}{\mathcal B}
\newcommand{\cD}{\mathcal D}
\newcommand{\cL}{\mathcal L}
\newcommand{\p}{\boldsymbol p}

\title{Global structure and holonomy of conserved resolutions in the cylindrical Dirac doublet}

\author[a]{Zhongze Guo,}
\author[b]{Bei Xu,}
\author[a]{Qiang Gu}

\affiliation[a]{Department of Physics and Institute of Theoretical Physics, University of Science and Technology Beijing, Beijing 100083, China}
\affiliation[b]{Institute for Advanced Study, Tsinghua University, Beijing 100084, China}

\emailAdd{guozhongze007@gmail.com}
\emailAdd{xubei0903@163.com}
\emailAdd{qgu@ustb.edu.cn}

\abstract{Cylindrical Dirac modes underlie constructions in rotating QCD matter, boost-invariant Dirac-field quantization in heavy-ion physics, and high-energy twisted-particle scattering. The corresponding complete spinor frames can be regarded as alternative bases, but their equivalence does not determine the global behavior of eigenlines selected by conserved observables. Within the positive-energy doublet of the free massive Dirac Hamiltonian, we compare three conserved resolutions: $K$, which couples spin to transverse momentum; $K_m$, a mass-dependent operator derived from the transverse Dirac Hamiltonian; and helicity. On the common regular domain away from the momentum axis, explicit smooth, single-valued $SU(2)$ transformations relate all three splittings. Although globally $SU(2)$-equivalent on this common domain, they exhibit three distinct global extension behaviors. The $K$ projectors have azimuth-dependent polar limits and do not extend continuously to the axis. For nonzero mass, the $K_m$ projectors extend smoothly over the enclosed momentum ball and define Chern-trivial eigenlines. The helicity projectors are smooth on every nonzero momentum sphere, but their eigenlines carry opposite unit Chern numbers and cannot extend through the enclosed origin. The parent positive-energy Dirac connection Abelianizes exactly in the $K$ eigenlines on fixed-azimuth meridians, whereas its full three-dimensional curvature has noncommuting components. We obtain the azimuthal Wilson loop in closed form and derive the exact conversion probability between the two $K$ branches under purely geometric positive-energy transport.
}


\begin{document}
\maketitle
\flushbottom

\section{Introduction}

Cylindrical solutions of the Dirac equation furnish mode functions for several distinct high-energy and field-theoretic settings. A transverse-helicity basis has been used in studies of chiral condensation and color superconductivity in globally rotating QCD matter~\cite{JiangLiao2016}; a different transverse separation underlies canonical Dirac-field quantization in longitudinally boost-invariant heavy-ion kinematics motivated by quark and antiquark production~\cite{Mihaila2006,Mihaila2009PRD}; and helicity-resolved Dirac--Bessel states serve as relativistic vortex states in high-energy twisted-particle scattering~\cite{Serbo2015TwistedScattering,Bliokh2017Review,Karlovets2017JHEP}. In each case the one-particle solutions are the mode functions from which the field or scattering states are constructed. Our analysis concerns the free bulk spinor resolutions entering these constructions; boundaries, spatially varying backgrounds, and Bogoliubov particle-production dynamics may impose additional domain or mixing constraints. Within this free bulk sector, the constructions are naturally organized as alternative resolutions of the same Dirac dynamics.

The familiar canonical-spin/helicity relation illustrates why a basis-level description is natural. At fixed momentum, the corresponding positive-energy spinors are connected by a momentum-dependent spin-$\tfrac12$ rotation~\cite{Wigner1939,Polyzou2013,Bliokh2011,Bliokh2017Review} and provide two frames of the same rank-two eigenspace. A passive transformation $U(\p)\to U(\p)G(\p)$ leaves the parent projector $P_+(\p)=U(\p)U^\dagger(\p)$ invariant, whereas fixing an eigenvalue of a conserved Hermitian operator selects a rank-one projector $\Pi_{A,s}(\p)$. The usual ``different-basis'' statement is therefore complete for the unresolved doublet but incomplete for its observable-selected eigenlines. Whether the corresponding projector families can be related
smoothly over a finite momentum-space domain is then a global
question.

A companion analysis~\cite{GuoXuGu2026Companion} establishes the common two-dimensional positive-energy solution space underlying the standard spin-polarized, transverse-separation, transverse-helicity, and helicity cylindrical constructions. The present work addresses the corresponding global extension problem. On the common regular domain
\begin{equation}
	M=\{\p:\kappa>0\},
	\label{eq:commonM}
\end{equation}
the three conserved splittings generated by $K$, $K_m$, and helicity are simultaneously defined. We show that their relation is stronger than pointwise $SU(2)$ equivalence: explicit smooth, single-valued $SU(2)$ maps intertwine the three ordered splittings throughout $M$. The nontrivial question is therefore whether these equivalences survive when the momentum-space loci excluded from $M$ are restored.

The cylindrical sector makes this question concrete because the three resolutions are not abstract choices of internal axis. The transverse-helicity operator $K=\beta(\boldsymbol\Sigma\times\boldsymbol p)_z$ is associated with the transverse-helicity construction used in rotating fermionic systems; the eigenspaces of a second conserved operator $K_m$ reproduce, up to representation and phase conventions, the transverse-separation sectors used in longitudinally boost-invariant Dirac-field quantization; and helicity gives the resolution used for relativistic Dirac--Bessel scattering states. All three act within the same positive-energy doublet on $M$, but the loci excluded from that common domain have different consequences for their continuations.

The resulting classification concerns global extension properties rather than the splittings on their common regular domain. The $K$ projectors have azimuth-dependent limits at $\kappa=0$ and therefore admit no continuous extension to the poles of a nonzero momentum sphere. For $m>0$, the $K_m$ resolving operator remains nondegenerate and its projectors extend smoothly through the entire enclosed momentum ball, implying Chern-trivial boundary eigenlines. Helicity is smooth on every nonzero momentum sphere and carries opposite unit Chern numbers, so its eigenlines cannot be extended through the enclosed origin. The three splittings are thus globally equivalent on $M$ but exhibit distinct global extension properties: nonextendable, smoothly extendable and Chern trivial, and smoothly extendable on each nonzero momentum sphere but topologically obstructed from extension through the enclosed origin.

The established non-Abelian Berry geometry of the positive-energy massive Dirac doublet~\cite{ShankarMathur1994,ChenPangPuWang2014,PuYamamoto2018} provides a complementary geometric structure. On a fixed-azimuth momentum meridian, the pulled-back connection and curvature are aligned with the projected $K$ direction, yielding exact Abelianization in the two $K$ eigenlines. Restoring the azimuthal direction produces noncommuting curvature components for $m>0$, and the corresponding azimuthal Wilson-loop spectrum can be evaluated in closed form. These results relate the symmetry-resolved extensions to the parent rank-two geometry without making the known parent connection the premise of the classification.

Observable-resolved subbundle topology and the non-Abelian Berry geometry of massive Dirac bands both have established precedents~\cite{Prodan2009,PalmerducaQinPhoton2024,PalmerducaQin2025,ShankarMathur1994,ChenPangPuWang2014,PuYamamoto2018}. The result specific to the present work is that the three conserved resolutions underlying standard cylindrical Dirac constructions in the high-energy settings above realize three distinct extension outcomes within the same exact positive-energy doublet. We construct their explicit smooth, single-valued $SU(2)$ intertwiners on the common regular domain and relate the resulting extension classification to exact meridional Abelianization and finite Wilson-loop holonomy.

The paper is organized as follows. Section~\ref{sec:doublet} formulates the common rank-two doublet, its conserved resolutions, and their global $SU(2)$ equivalence on $M$. Section~\ref{sec:geometry} develops their quantum geometry and relation to the parent non-Abelian connection. Section~\ref{sec:holonomy} derives the exact azimuthal Wilson loop. Section~\ref{sec:global} classifies the distinct global extension properties, and Section~\ref{sec:discussion} summarizes the resulting hierarchy and its physical scope.

\section{Cylindrical Dirac doublet and symmetry reductions}
\label{sec:doublet}

\subsection{Rank-two sector and projective fiber}
\label{subsec:projectivefiber}

We work in natural units $\hbar=c=1$, take $m>0$, and write the free Dirac Hamiltonian as
\begin{equation}
H(\p)=\boldsymbol\alpha\cdot\p+\beta m,
\qquad
E=\sqrt{m^2+\p^2}>0.
\label{eq:H}
\end{equation}
With $\boldsymbol\Sigma=\operatorname{diag}(\boldsymbol\sigma,\boldsymbol\sigma)$ and $J_z=L_z+\Sigma_z/2$, the positive-energy projector is
\begin{equation}
P_+(\p)=\frac12\left(\one+\frac{H(\p)}{E}\right),
\qquad
\cE^{(+)}_{\p}=\Ran P_+(\p)\simeq\mathbb C^2.
\label{eq:Pplus}
\end{equation}
The collection $\cE^{(+)}\equiv\bigsqcup_{\p}\cE^{(+)}_{\p}$ defines the rank-two positive-energy bundle over momentum space. At fixed $\p$, a change of frame acts within $\cE^{(+)}_{\p}$ and leaves $P_+$ invariant, whereas an additional conserved Hermitian observable may resolve this fiber into rank-one spectral subspaces.

For cylindrical kinematics we parameterize momentum by
\begin{equation}
\p=\kappa\,\boldsymbol e_r(\phi)+k_z\boldsymbol e_z,
\qquad
p=|\p|=\sqrt{\kappa^2+k_z^2},
\qquad \kappa\geq0.
\label{eq:cylp}
\end{equation}
The azimuth $\phi$ is a coordinate on the parent momentum-space bundle and should be distinguished from the real-space azimuth entering a cylindrical Bessel mode. At fixed on-shell $(\kappa,k_z)$ and $J_z=n+\tfrac12$, the regular positive-energy cylindrical solutions span a two-dimensional multiplicity space,
\begin{equation}
\cD^{(+)}_{n\kappa k_z}\simeq\mathbb C^2.
\label{eq:doublet}
\end{equation}
For $\kappa>0$ and $n\in\mathbb Z$, this multiplicity space is related explicitly to the internal positive-energy fibers of the plane-wave components entering the cylindrical Fourier synthesis. Writing
\begin{equation}
\p(\phi)=\kappa\,\boldsymbol e_r(\phi)+k_z\boldsymbol e_z,
\qquad
G_B(\phi)=
\begin{pmatrix}
1&0\\
0&-\ii e^{\ii\phi}
\end{pmatrix},
\label{eq:GB}
\end{equation}
and $c=(c_\uparrow,c_\downarrow)^T\in\mathbb C^2$, the corresponding regular cylindrical mode may be written, up to a common normalization, as
\begin{equation}
\Psi_c(r,\varphi,z,t)
\propto
 e^{-\ii Et+\ii k_z z}\,\ii^{-n}
\int_0^{2\pi}\frac{\dd\phi}{2\pi}
 e^{\ii n\phi+\ii\kappa r\cos(\phi-\varphi)}
 U\bigl(\p(\phi)\bigr)G_B(\phi)c .
\label{eq:BesselFourier}
\end{equation}
Here $\phi$ is the momentum azimuth integrated around the cone, whereas $\varphi$ is the real-space azimuth of the cylindrical wave. Equation~\eqref{eq:BesselFourier} follows from
\begin{equation}
\int_0^{2\pi}\frac{\dd\phi}{2\pi}
 e^{\ii\ell\phi+\ii\kappa r\cos(\phi-\varphi)}
 =\ii^{\ell}e^{\ii\ell\varphi}J_\ell(\kappa r),
\label{eq:BesselIdentity}
\end{equation}
and shows explicitly how the constant coefficient vector $c$ parametrizes the two-dimensional Bessel multiplicity while the internal spinor at each momentum component lies in the plane-wave fiber $\cE^{(+)}_{\p(\phi)}$. The matrix $G_B(\phi)$ is the unitary change of internal frame appropriate to the cylindrical Fourier synthesis; it does not introduce an additional physical reduction.

The momentum-space bundle studied below is therefore the internal positive-energy bundle of the plane-wave components entering Eq.~\eqref{eq:BesselFourier}. All extension and Chern statements refer to the momentum-dependent projectors $\Pi_{A,s}(\p)$ in these internal fibers. They do not assert that every fixed-$n$ Bessel basis vector, or every normalized spatial wave packet constructed from it, possesses the same nonsingular extension to the momentum axis. Indeed, at $\kappa=0$ the Fourier cone collapses and a particular fixed-angular-momentum Bessel frame may lose rank even when the underlying internal momentum projector has a regular limit. An explicit realization of Eq.~\eqref{eq:BesselFourier} is given in Sec.~\ref{sec:explicitdoublet}.

The complete regular cylindrical family obtained in Ref.~\cite{GuoXuGu2026Companion} provides a convenient local coordinate on the projectivized multiplicity space. In a conventional spin-polarized cylindrical frame,
\begin{equation}
\Psi_\lambda\propto \Psi_\uparrow+z(\lambda)\Psi_\downarrow,
\qquad
z(\lambda)=-\frac{\ii}{\kappa}\left(k_z-\frac{E+m}{\lambda}\right),
\label{eq:lambdamap}
\end{equation}
with inverse
\begin{equation}
\lambda=\frac{E+m}{k_z-\ii\kappa z}.
\label{eq:lambdainverse}
\end{equation}
After quotienting by nonzero complex normalization, the physical rays form
\begin{equation}
\mathbb P(\cD^{(+)})\simeq\mathbb CP^1.
\end{equation}
Accordingly, $z$, or equivalently $\lambda$ on this chart, parametrizes a ray within the fiber rather than a point of the momentum-space base. The coordinate $\lambda$ is frame dependent; the invariant object associated with a physical state is its rank-one projector. A conserved reduction therefore determines momentum-dependent projective sections $\lambda_{A,\pm}(\p)$ wherever both the spectral splitting and this coordinate chart are regular.

For the momentum-space geometry we employ the globally smooth canonical positive-energy frame
\begin{equation}
U(\p)=
\sqrt{\frac{E+m}{2E}}
\begin{pmatrix}
\one_2\\[1mm]
\dfrac{\boldsymbol\sigma\cdot\p}{E+m}
\end{pmatrix},
\qquad
U^\dagger U=\one_2,
\qquad
UU^\dagger=P_+.
\label{eq:canonicalframe}
\end{equation}
Let $\Gamma_A$ be a Hermitian involution satisfying $[\Gamma_A,P_+]=0$ and having one eigenvalue of each sign within the positive-energy doublet, equivalently $\Tr_2(U^\dagger\Gamma_AU)=0$. Its reduced representation and rank-one spectral projectors are
\begin{equation}
\gamma_A=U^\dagger\Gamma_AU,
\qquad
\pi_{A,s}=\frac12(\one+s\gamma_A),
\qquad
\Pi_{A,s}=U\pi_{A,s}U^\dagger
=P_+\frac{\one+s\Gamma_A}{2},
\quad s=\pm1.
\label{eq:physicalprojector}
\end{equation}
Under a frame transformation $U\to UG$, with $G(\p)\in U(2)$, one has $\gamma_A\to G^\dagger\gamma_A G$ and $\pi_{A,s}\to G^\dagger\pi_{A,s}G$, whereas the four-component projector $\Pi_{A,s}$ is unchanged. The distinction between a passive frame transformation and a change of observable-selected eigenspace is therefore expressed directly at the projector level.

We finally introduce the cylindrical Pauli matrices
\begin{equation}
\sigma_r=\cos\phi\,\sigma_x+\sin\phi\,\sigma_y,
\qquad
\sigma_\phi=-\sin\phi\,\sigma_x+\cos\phi\,\sigma_y,
\label{eq:cylpauli}
\end{equation}
which will be used to represent the conserved axes inside the positive-energy fiber.

\subsection{Transverse conserved integrals and the internal Pauli algebra}

The first conserved resolution is associated with transverse helicity.  Balantekin and DeWeerd formulated a transverse-helicity member of a cylindrical complete set of commuting observables~\cite{BalantekinDeWeerd1995}; published applications subsequently used the corresponding quantum number and eigenmodes in rotating fermionic matter and confined Dirac systems~\cite{JiangLiao2016,Khosravi2019PRB}.  We use the equivalent unnormalized conserved integral
\begin{equation}
K=\beta(\Sigma_xp_y-\Sigma_yp_x)
=\beta(\boldsymbol\Sigma\times\p)_z,
\label{eq:Kdef}
\end{equation}
with the overall sign convention adopted in Ref.~\cite{GuoXuGu2026Companion}.  On the regular domain $\kappa>0$, Eq.~\eqref{eq:Kdef} is algebraically equivalent, up to normalization and convention-dependent signs, to the previously used transverse-helicity operator.  The spin--momentum form is structurally useful here because it exposes a transverse Rashba-type coupling and makes the projection and anticommutation algebra immediate:
\begin{equation}
[K,H]=[K,p_z]=[K,J_z]=0,
\qquad
K^2=p_\perp^2.
\label{eq:Kalg}
\end{equation}
Hence, for $\kappa>0$,
\begin{equation}
\Gamma_K=\frac{K}{\kappa},
\qquad
\Gamma_K^2=\one.
\label{eq:GammaK}
\end{equation}
On the momentum axis the two eigenvalues $\pm\kappa$ coalesce, so $K$ no longer selects two branches uniquely.  This degeneracy alone does not imply nonextendability; the actual obstruction is established in Sec.~\ref{subsec:Kglobal} by the azimuth-dependent limits of the off-axis spectral projectors.

The second conserved resolution is generated by
\begin{equation}
K_m=\ii\beta(\alpha_xp_y-\alpha_yp_x)-m\Sigma_z.
\label{eq:Kmdef}
\end{equation}
This operator is not introduced as an ad hoc mass regularization of $K$.  It follows directly from the transverse Dirac operator
\begin{equation}
H_\perp=\alpha_xp_x+\alpha_yp_y+\beta m,
\end{equation}
through the factorization
\begin{equation}
K_m=-\beta\Sigma_z H_\perp.
\label{eq:KmHperp}
\end{equation}
Its eigenspaces reproduce the transverse-separation sectors used in the longitudinally boost-invariant cylindrical quantization of Mihaila, Dawson, and Cooper (MDC)~\cite{Mihaila2006}.  The identification can be checked directly.  In the MDC convention let
\begin{equation}
B=\ii\gamma^1_{\rm MDC}\gamma^2_{\rm MDC},
\qquad
K_2=(\gamma^1_{\rm MDC}\partial_x+\gamma^2_{\rm MDC}\partial_y+m)B,
\end{equation}
and use $p_j=-\ii\partial_j$ together with
\begin{equation}
\ii\gamma^\mu_{\rm MDC}=S\gamma_D^\mu S^\dagger,
\qquad
S=\frac{1}{\sqrt2}
\begin{pmatrix}
\ii\one_2&-\ii\one_2\\
\one_2&\one_2
\end{pmatrix}.
\end{equation}
Then
\begin{equation}
S^\dagger B K_2 B S=-\beta\Sigma_zH_\perp=K_m.
\label{eq:MDCmapping}
\end{equation}
Thus Eq.~\eqref{eq:KmHperp} is not merely a qualitative analogy: it is the Dirac-representation form of the MDC transverse-separation operator, up to the stated convention transformation.

For completeness, the relevant algebra can be verified directly.  Define
\begin{equation}
A=\Sigma_xp_y-\Sigma_yp_x,
\qquad
T=\alpha_xp_y-\alpha_yp_x,
\end{equation}
so that $K=\beta A$ and $K_m=\ii\beta T-m\Sigma_z$.  The Dirac algebra gives
\begin{equation}
A^2=T^2=p_\perp^2,
\qquad
[A,T]=0,
\qquad
\{A,\Sigma_z\}=0,
\qquad
\{\ii\beta T,\Sigma_z\}=0.
\label{eq:ATids}
\end{equation}
Consequently,
\begin{equation}
K_m^2=m^2+p_\perp^2,
\qquad
\{K,K_m\}=0.
\label{eq:KmKanti}
\end{equation}
The conservation of $K_m$ follows directly from Eq.~\eqref{eq:KmHperp}.  The operator $-\beta\Sigma_z$ commutes with $H_\perp$, whereas both $-\beta\Sigma_z$ and $H_\perp$ anticommute with $\alpha_z$; their product therefore commutes with the longitudinal term $\alpha_z p_z$.  Together with axial rotational covariance and the absence of $z$ dependence, this yields
\begin{equation}
[K_m,H]=[K_m,p_z]=[K_m,J_z]=0.
\label{eq:Kmalg}
\end{equation}
Unlike the transverse-helicity spectrum, the $K_m$ spectrum remains separated on the momentum axis for $m\neq0$.

Introducing
\begin{equation}
\Lambda=\sqrt{m^2+\kappa^2},
\qquad
\Gamma_m=\frac{K_m}{\Lambda},
\end{equation}
Eqs.~\eqref{eq:KmKanti} and~\eqref{eq:GammaK} imply
\begin{equation}
\Gamma_m^2=\one,
\qquad
\{\Gamma_K,\Gamma_m\}=0.
\label{eq:anti}
\end{equation}
The two normalized conserved axes therefore generate a Pauli algebra on each regular multiplicity space.  With $\Gamma_1=\Gamma_K$, $\Gamma_2=\Gamma_m$, and
\begin{equation}
\Gamma_3=-\ii\Gamma_K\Gamma_m,
\end{equation}
one obtains
\begin{equation}
\Gamma_a\Gamma_b=\delta_{ab}\one+\ii\epsilon_{abc}\Gamma_c,
\qquad
[\Gamma_a,\Gamma_b]=2\ii\epsilon_{abc}\Gamma_c.
\label{eq:paulialg}
\end{equation}
This algebra acts on the residual positive-energy multiplicity and should not be interpreted as an additional spacetime $SU(2)$ symmetry.  Its significance is internal: it places the $K$ and $K_m$ resolutions on two orthogonal axes of the same two-dimensional fiber and prepares the comparison with helicity in the following subsection.

\subsection{Local symmetry axes inside the positive-energy fiber}

The conserved operators acquire a particularly transparent form after restriction to the positive-energy doublet.  For a normalized operator $\Gamma_A$ preserving this sector, define
\begin{equation}
\gamma_A(\p)=U^\dagger(\p)\Gamma_A(\p)U(\p),
\label{eq:gammaAdef}
\end{equation}
as in Eq.~\eqref{eq:physicalprojector}.  Each $\gamma_A$ is a traceless Hermitian $2\times2$ matrix with $\gamma_A^2=\one$ and therefore determines a unit vector on the internal Bloch sphere.  For the transverse-helicity integral,
\begin{equation}
\gamma_K=-\sigma_\phi,
\label{eq:gK}
\end{equation}
whereas the transverse-separation integral $K_m$ gives
\begin{equation}
\gamma_m=n_m^{r}\sigma_r+n_m^{z}\sigma_z,
\qquad
n_m^{r}=\frac{\kappa k_z}{\Lambda(E+m)},
\qquad
n_m^{z}=-\frac{m(E+m)+\kappa^2}{\Lambda(E+m)},
\label{eq:gm}
\end{equation}
with $(n_m^{r})^2+(n_m^{z})^2=1$.  Equation~\eqref{eq:gm} is the internal axis associated with the two $K_m$ eigenspaces of eigenvalues $s\Lambda$, $s=\pm1$.

For $p>0$, helicity provides a third conserved resolution,
\begin{equation}
\Gamma_h=\frac{\boldsymbol\Sigma\cdot\p}{p},
\qquad
[\Gamma_h,H]=0,
\qquad
\Gamma_h^2=\one,
\label{eq:Gammah}
\end{equation}
whose positive-energy representation is
\begin{equation}
\gamma_h=\frac{\kappa}{p}\sigma_r+\frac{k_z}{p}\sigma_z.
\label{eq:gh}
\end{equation}
Thus $K$, $K_m$, and helicity define three momentum-dependent axes in the same two-dimensional fiber.  Their relative orientations satisfy
\begin{equation}
\{\gamma_K,\gamma_m\}=\{\gamma_K,\gamma_h\}=0,
\qquad
\frac12\{\gamma_m,\gamma_h\}
=-\frac{mk_z}{\Lambda p}\one_2.
\label{eq:relativeaxes}
\end{equation}
The $K$ axis is orthogonal to both meridional axes, while the relative angle between the $K_m$ and helicity axes depends on $m$ and momentum.

The anticommutation relations in Eq.~\eqref{eq:relativeaxes} imply a stronger statement than pointwise equivalence. On the common regular domain $M$ of Eq.~\eqref{eq:commonM}, define, for $B=m,h$,
\begin{equation}
g_{B\leftarrow K}(\p)
=\frac{\one_2+\gamma_B(\p)\gamma_K(\p)}{\sqrt2}.
\label{eq:gBK}
\end{equation}
Because $\gamma_B$ and $\gamma_K$ are Hermitian involutions and anticommute, $\gamma_B\gamma_K$ is traceless and anti-Hermitian with square $-\one_2$. Consequently
\begin{equation}
g_{B\leftarrow K}\in SU(2),
\qquad
g_{B\leftarrow K}\gamma_K g_{B\leftarrow K}^\dagger=\gamma_B,
\qquad
\pi_{B,s}=g_{B\leftarrow K}\pi_{K,s}g_{B\leftarrow K}^\dagger.
\label{eq:commonSU2}
\end{equation}
All three reduced axes are smooth and $2\pi$-periodic throughout $M$, so the maps~\eqref{eq:gBK} are smooth and single valued there. In particular,
\begin{equation}
g_{h\leftarrow m}
=g_{h\leftarrow K}g_{m\leftarrow K}^\dagger
\label{eq:ghm}
\end{equation}
provides an explicit global equivalence between the $K_m$ and helicity splittings on $M$. Thus the three ordered spectral decompositions are globally $SU(2)$-equivalent wherever they are simultaneously defined; the distinction studied below concerns the inequivalent ways in which these decompositions extend beyond that common regular domain.

The familiar canonical-spin/helicity relation provides a useful reference for this distinction.  In the canonical frame~\eqref{eq:canonicalframe}, let $\gamma_c=\sigma_z$.  This axis may be viewed not only as a frame label but also as the positive-energy restriction of a conserved canonical (Foldy--Wouthuysen) spin observable~\cite{Polyzou2013,Bliokh2011}, with projector $\Pi_{c,s}=U(\one+s\sigma_z)U^\dagger/2$.  Its eigenlines are globally trivial for $m>0$ and therefore already furnish the standard trivial counterpart to the nontrivial helicity lines.  We do not include canonical spin in the three-way classification below because our selection criterion is the three conserved resolutions tied directly to the cylindrical constructions introduced above.  Writing
\begin{equation}
\cos\vartheta=\frac{k_z}{p},
\qquad
\sin\vartheta=\frac{\kappa}{p},
\end{equation}
the standard spin-$\tfrac12$ canonical-to-helicity spin rotation takes the form
\begin{equation}
D_{c\to h}(\p)
=\exp\!\left(\frac{\ii\vartheta}{2}\gamma_K\right),
\qquad
\gamma_h=D_{c\to h}\,\gamma_c\,D_{c\to h}^\dagger.
\label{eq:wignercheck}
\end{equation}
Here Eq.~\eqref{eq:gK} shows that the projected transverse-helicity operator supplies precisely the internal rotation axis connecting the canonical-spin direction to helicity.  The canonical-to-helicity spin rotation is standard~\cite{Wigner1939,Polyzou2013}; its role here is diagnostic.  In particular, the familiar canonical/helicity contrast already shows that observable-selected eigenlines can have different global topology.  Our new comparison is therefore not the abstract existence of such a contrast, but the exact relation among the three conserved cylindrical resolutions, their distinct global extension properties, and their transport under the parent connection.

A frame-independent measure of the pointwise relation between two reductions follows directly from their spectral projectors.  If
\begin{equation}
\gamma_A=\boldsymbol n_A\cdot\boldsymbol\sigma,
\qquad
\gamma_B=\boldsymbol n_B\cdot\boldsymbol\sigma,
\end{equation}
then
\begin{equation}
\Tr(\pi_{A,s}\pi_{B,t})
=\frac12\left(1+st\,\boldsymbol n_A\cdot\boldsymbol n_B\right).
\label{eq:overlap}
\end{equation}
Because $\boldsymbol n_K$ is orthogonal to both $\boldsymbol n_m$ and $\boldsymbol n_h$, each $K$ eigenstate has equal projection probability onto the two $K_m$ eigenspaces and likewise onto the two helicity eigenspaces at fixed momentum.  These pointwise overlap relations characterize the relative orientations of the conserved axes within the common domain; they do not constrain the extensions of the corresponding projectors across the excluded loci.

\subsection{Exact cylindrical realization of the projector sections}
\label{sec:explicitdoublet}

Taking $c=(1,0)^T$ and $c=(0,1)^T$ in the Fourier synthesis~\eqref{eq:BesselFourier} yields the regular multiplicity frame constructed in Ref.~\cite{GuoXuGu2026Companion}.  After suppressing the common factor $e^{-\ii E t+\ii k_z z+\ii n\varphi}$,
\begin{equation}
\Psi_\uparrow=\mathcal N
\begin{pmatrix}
J_n\\
0\\
\dfrac{k_z}{E+m}J_n\\
\dfrac{\ii\kappa}{E+m}e^{\ii\varphi}J_{n+1}
\end{pmatrix},
\qquad
\Psi_\downarrow=\mathcal N
\begin{pmatrix}
0\\
e^{\ii\varphi}J_{n+1}\\
-\dfrac{\ii\kappa}{E+m}J_n\\
-\dfrac{k_z}{E+m}e^{\ii\varphi}J_{n+1}
\end{pmatrix},
\label{eq:Besselbasis}
\end{equation}
where $J_\ell\equiv J_\ell(\kappa r)$ and $\varphi$ is the real-space azimuth.  The common factor $\mathcal N$ denotes the usual generalized normalization of ideal Bessel modes (equivalently, the coefficient-space normalization inherited from the plane-wave Fourier synthesis), not a pointwise normalization in real space.  As for plane waves, an ideal Bessel mode is delta-normalized; a normalizable beam requires a momentum-space envelope.  Equation~\eqref{eq:Besselbasis} is a local frame of $\cD^{(+)}_{n\kappa k_z}$ rather than an additional physical reduction.

Because Eq.~\eqref{eq:BesselFourier} relates the Bessel coefficient space to the canonical plane-wave frame through $G_B(\phi)$, the reduced matrices in the ordered coefficient basis $(\Psi_\uparrow,\Psi_\downarrow)$ are obtained by conjugation with $G_B$. With $\tau_i$ denoting Pauli matrices on this coefficient space, one finds
\begin{equation}
G_B^\dagger\gamma_KG_B=\tau_x,
\qquad
G_B^\dagger\gamma_mG_B=b_m\tau_y-a_m\tau_z,
\qquad
G_B^\dagger\gamma_hG_B=\frac{\kappa}{p}\tau_y+\frac{k_z}{p}\tau_z.
\label{eq:GBconjugation}
\end{equation}
Equivalently, the three normalized conserved observables reduce to
\begin{equation}
\gamma_K=\tau_x,
\qquad
\gamma_m=b_m\tau_y-a_m\tau_z,
\qquad
\gamma_h=\frac{\kappa}{p}\tau_y+\frac{k_z}{p}\tau_z,
\label{eq:Besselreduced}
\end{equation}
where
\begin{equation}
 a_m=\frac{m(E+m)+\kappa^2}{\Lambda(E+m)},
 \qquad
 b_m=\frac{\kappa k_z}{\Lambda(E+m)},
 \qquad
 a_m^2+b_m^2=1.
\label{eq:ambm}
\end{equation}
Hence $K$, $K_m$, and helicity define three pairs of spectral lines within the same exact Bessel doublet.

Using the projective coordinate of Sec.~\ref{subsec:projectivefiber}, the corresponding eigenline conditions give
\begin{align}
\lambda_{K,\eta}
&=\frac{E+m}{k_z-\ii\eta\kappa},
&& \eta=\pm1,
\label{eq:lambdaK}\\
\lambda_{m,s}
&=\frac{k_z}{E+s\Lambda}
 =\frac{E-s\Lambda}{k_z},
&& s=\pm1,
\label{eq:lambdam}\\
\lambda_{h,s}
&=s\frac{E+m}{p},
&& s=\pm1.
\label{eq:lambdah}
\end{align}
The two algebraically equivalent forms in Eq.~\eqref{eq:lambdam} are useful on complementary kinematic patches. At $k_z=0$ the two $K_m$ lines are represented without ambiguity by homogeneous coordinates,
\begin{equation}
[\lambda_{m,+}:1]=[k_z:E+\Lambda],
\qquad
[\lambda_{m,-}:1]=[E+\Lambda:k_z],
\label{eq:lambdamhomogeneous}
\end{equation}
so that the limiting affine coordinates are respectively $0$ and $\infty$. More generally, an affine divergence is removed by passing to the reciprocal projective chart and does not signal a singular physical projector. This chart statement applies only for $\kappa>0$; at $\kappa=0$ the cylindrical coordinate relation~\eqref{eq:lambdamap} itself degenerates with the Fourier cone and must not be interpreted as a projective-chart singularity. The $K_m$ sections coincide with the transverse-separation sectors of Ref.~\cite{Mihaila2006}.  Equations~\eqref{eq:Besselreduced}--\eqref{eq:lambdah} therefore provide the required exact-mode realization of the three projector families; their global extendibility and topology must be determined from the projectors themselves.

\section{Parent non-Abelian geometry and meridional reduction}
\label{sec:geometry}

\subsection{Positive-energy connection and curvature}

In the global canonical frame~\eqref{eq:canonicalframe}, the Berry connection induced by the positive-energy projector is represented by the matrix-valued Wilczek--Zee connection~\cite{Wilczek1984}
\begin{equation}
\cA=\ii U^\dagger\dd U.
\label{eq:Adef}
\end{equation}
For the canonical frame, direct evaluation gives the standard positive-energy Dirac connection~\cite{ShankarMathur1994,ChenPangPuWang2014,PuYamamoto2018},
\begin{equation}
\cA=-\frac{(\p\times\dd\p)\cdot\boldsymbol\sigma}{2E(E+m)}.
\label{eq:Acompact}
\end{equation}
It is traceless in this frame.  More generally, under a momentum-dependent change of positive-energy frame $U\to UG(\p)$ with $G\in U(2)$,
\begin{equation}
\cA\to G^\dagger\cA G+\ii G^\dagger\dd G,
\end{equation}
whereas the curvature
\begin{equation}
\cF=\dd\cA-\ii\cA\wedge\cA
\end{equation}
transforms covariantly, $\cF\to G^\dagger\cF G$.  In the canonical frame its Cartesian components are
\begin{equation}
\cF_{ij}=-\frac{\epsilon_{ijk}}{2E^3}
\left[m\sigma_k+\frac{p_k(\p\cdot\boldsymbol\sigma)}{E+m}\right],
\label{eq:Fcart}
\end{equation}
so that $\Tr\cF_{ij}=0$.  Equations~\eqref{eq:Acompact} and~\eqref{eq:Fcart} are established properties of the massive positive-energy Dirac bundle and are included here to fix conventions and to provide the parent geometry for the symmetry-resolved reductions below.

For $m>0$, the frame~\eqref{eq:canonicalframe} is smooth on all of $\mathbb R^3_{\p}$, including $\p=0$.  The parent rank-two bundle is therefore globally trivial, notwithstanding its nonvanishing matrix curvature.  In this sense, the ``$SU(2)$ geometry'' used below refers to the traceless non-Abelian sector of the connection rather than to a nontrivial topology of the parent bundle.  This separation between parent-bundle topology and connection geometry is essential for the comparison with the rank-one eigenbundles in Sec.~\ref{sec:global}.

\subsection{Cylindrical pullback and exact meridional Abelianization}
\label{subsec:meridional}

Using Eq.~\eqref{eq:cylp},
\begin{equation}
\dd\p=\boldsymbol e_r\dd\kappa+\kappa\boldsymbol e_\phi\dd\phi+\boldsymbol e_z\dd k_z,
\end{equation}
the parent connection~\eqref{eq:Acompact} takes the cylindrical form
\begin{align}
\cA_\kappa&=-\frac{k_z}{2E(E+m)}\sigma_\phi,
\label{eq:Akappa}\\
\cA_{k_z}&=\frac{\kappa}{2E(E+m)}\sigma_\phi,
\label{eq:Akz}\\
\cA_\phi&=\frac{\kappa k_z\sigma_r-\kappa^2\sigma_z}{2E(E+m)}.
\label{eq:Aphi}
\end{align}
The corresponding curvature components follow from Eq.~\eqref{eq:Fcart}.  In particular,
\begin{equation}
\cF_{\kappa k_z}
=\frac{m}{2E^3}\sigma_\phi
=-\frac{m}{2E^3}\gamma_K,
\label{eq:FselectK}
\end{equation}
while
\begin{align}
\cF_{\kappa\phi}&=-\frac{\kappa}{2E^3}
\left[
\frac{\kappa k_z}{E+m}\sigma_r+
\left(m+\frac{k_z^2}{E+m}\right)\sigma_z
\right],
\label{eq:Fkphi}\\
\cF_{\phi k_z}&=-\frac{\kappa}{2E^3}
\left[
\left(m+\frac{\kappa^2}{E+m}\right)\sigma_r+
\frac{\kappa k_z}{E+m}\sigma_z
\right].
\label{eq:Fphikz}
\end{align}
Equation~\eqref{eq:FselectK} is a direct consistency relation between two independently defined structures: \(\gamma_K\) is obtained by projecting the conserved transverse operator \(K\), whereas \(\cF_{\kappa k_z}\) follows from the parent Wilczek--Zee connection.  On a fixed-azimuth meridian, the two select the same internal direction.

The coincidence extends from the curvature to the full pulled-back connection.  Setting \(\dd\phi=0\) gives
\begin{equation}
\cA\big|_{\phi=\mathrm{const.}}
=\frac{k_z\,\dd\kappa-\kappa\,\dd k_z}{2E(E+m)}\,\gamma_K,
\qquad
\gamma_K=-\sigma_\phi.
\label{eq:Ameridian}
\end{equation}
At fixed \(\phi\), \(\gamma_K\) is independent of \((\kappa,k_z)\) and commutes with the pulled-back connection.  It is therefore covariantly constant on the meridian, and the parent \(SU(2)\) connection reduces exactly to the \(U(1)\) subalgebra generated by \(\gamma_K\).  Consequently, parallel transport on this submanifold is diagonal in the \(K\) eigenlines and requires no non-Abelian path ordering.

If \(\gamma_K|K,\eta\rangle=\eta|K,\eta\rangle\), the induced Abelian connections are
\begin{equation}
\cA^{K,\eta}\big|_{\phi=\mathrm{const.}}
=\eta\,\frac{k_z\,\dd\kappa-\kappa\,\dd k_z}{2E(E+m)},
\label{eq:KbranchA}
\end{equation}
with Berry curvature
\begin{equation}
\Omega_{\kappa k_z}^{K,\eta}=-\eta\frac{m}{2E^3}.
\label{eq:Kcurv}
\end{equation}
This agrees exactly with the symmetry-resolved curvature obtained directly from the cylindrical \(K\) branches in Ref.~\cite{GuoXuGu2026Companion}.  Thus the branch geometry derived from the conserved-operator reduction is the eigenline reduction of the parent non-Abelian connection on the cylindrical meridian.  The reduction is intrinsically meridional: the azimuthal components in Eqs.~\eqref{eq:Aphi}, \eqref{eq:Fkphi}, and~\eqref{eq:Fphikz} involve additional Pauli directions and restore the non-Abelian structure examined below.

For purely geometric transport within the positive-energy doublet, after removal of the common dynamical phase, preservation of an instantaneous resolved line is equivalent to the covariant condition
\begin{equation}
D_t\pi_{A,s}=\dot p_iD_i\pi_{A,s}=0,
\qquad
D_i\pi_{A,s}=\partial_i\pi_{A,s}-\ii[\cA_i,\pi_{A,s}].
\label{eq:branchcovariant}
\end{equation}
For the \(K\) projectors on a fixed-azimuth meridian, Eq.~\eqref{eq:Ameridian} gives \(D_\kappa\pi_{K,\eta}=D_{k_z}\pi_{K,\eta}=0\).  Thus the exact meridional Abelianization also means that positive-energy geometric parallel transport preserves the two \(K\) eigenlines along such paths.

\subsection{Meridional quantum geometry of the three reductions}
\label{sec:qgtmain}

On a fixed-azimuth meridian, the three conserved resolutions are defined over the same two-parameter base $(\kappa,k_z)$ and may therefore be compared directly at the level of their physical projectors. For a rank-one projector $\Pi_{A,s}$, we use the projector representation of the quantum metric and Berry curvature~\cite{ProvostVallee1980,Berry1984,GrafPiechon2021},
\begin{equation}
g_{ij}^{A,s}=\frac12\Tr[(\partial_i\Pi_{A,s})(\partial_j\Pi_{A,s})],
\qquad
\Omega_{ij}^{A,s}=\ii\Tr\!\left\{\Pi_{A,s}[\partial_i\Pi_{A,s},\partial_j\Pi_{A,s}]\right\},
\label{eq:projectorQGT}
\end{equation}
with $i,j\in\{\kappa,k_z\}$.  These tensors describe the internal four-component Dirac projectors of the momentum components entering the Fourier synthesis.  They do not include additional parameter dependence of a radial or longitudinal envelope used to regularize an ideal Bessel mode into a normalizable wave packet.  In particular, the nonzero geometry below is not obtained by differentiating only the constant two-component Bessel coefficient vector; it also contains the momentum dependence of the embedding $U(\p)$ of the positive-energy subspace in the full Dirac space.

A useful gauge-covariant decomposition makes this separation explicit.  For the reduced rank-one projector $\pi_{A,s}$ define $D_i\pi_{A,s}=\partial_i\pi_{A,s}-\ii[\cA_i,\pi_{A,s}]$.  The physical eigenline curvature is
\begin{equation}
\Omega^{A,s}_{ij}
=\Tr_2(\pi_{A,s}\cF_{ij})
+\ii\Tr_2\!\left\{\pi_{A,s}[D_i\pi_{A,s},D_j\pi_{A,s}]\right\}.
\label{eq:covariantbranchcurv}
\end{equation}
Thus a branch curvature is not, in general, obtained by taking only an expectation value of the parent non-Abelian curvature.  Writing $\gamma_A=\boldsymbol n_A\cdot\boldsymbol\sigma$, we define the Bloch-vector covariant derivative by
\[
(D_i\boldsymbol n_A)\cdot\boldsymbol\sigma
\equiv D_i\gamma_A
=\partial_i\gamma_A-\ii[\cA_i,\gamma_A].
\]
The metric in the canonical Dirac frame then decomposes as
\begin{align}
g^{A,s}_{ij}&=q_{ij}+\frac14D_i\boldsymbol n_A\cdot D_j\boldsymbol n_A,
\label{eq:metricdecomp}\\
q_{ij}&=\frac{1}{4E^2}\left[
\partial_i\p\cdot\partial_j\p
-\frac{(\p\cdot\partial_i\p)(\p\cdot\partial_j\p)}{E^2}
\right].
\label{eq:qparent}
\end{align}
On a fixed-azimuth meridian,
\begin{equation}
q=\frac{1}{4E^4}
\begin{pmatrix}
m^2+k_z^2&-\kappa k_z\\
-\kappa k_z&m^2+\kappa^2
\end{pmatrix},
\label{eq:qmeridian}
\end{equation}
and direct covariant differentiation gives
\begin{align}
g^K&=q,\label{eq:gKdecomp}\\
g^m&=q+\frac{m^2k_z^2}{4\Lambda^4E^2}
\begin{pmatrix}1&0\\0&0\end{pmatrix},\label{eq:gmdecomp}\\
g^h&=q+\frac{m^2}{4E^2p^4}
\begin{pmatrix}k_z^2&-\kappa k_z\\-\kappa k_z&\kappa^2\end{pmatrix}.
\label{eq:ghdecomp}
\end{align}
Equations~\eqref{eq:covariantbranchcurv}--\eqref{eq:ghdecomp} provide a compact derivation of the branch-resolved tensors listed below.  The quantum metric is identical for the two branches of each reduction, whereas the Berry curvature is branch odd whenever it is nonzero.

For the $K$ reduction, the companion analysis~\cite{GuoXuGu2026Companion} gives, in the present notation,
\begin{subequations}\label{eq:QGTK}
\begin{align}
g^{K}_{\kappa\kappa}&=\frac{m^2+k_z^2}{4E^4},
&g^{K}_{k_zk_z}&=\frac{m^2+\kappa^2}{4E^4},\\
g^{K}_{\kappa k_z}&=-\frac{\kappa k_z}{4E^4},
&\Omega^{K,\eta}_{\kappa k_z}&=-\eta\frac{m}{2E^3}.
\end{align}
\end{subequations}
For the transverse-separation reduction $K_m$,
\begin{subequations}\label{eq:QGTm}
\begin{align}
g^{m}_{\kappa\kappa}&=\frac{m^2}{4\Lambda^4}+\frac{\kappa^2k_z^2}{4\Lambda^2E^4},
&g^{m}_{k_zk_z}&=\frac{\Lambda^2}{4E^4},\\
g^{m}_{\kappa k_z}&=-\frac{\kappa k_z}{4E^4},
&\Omega^{m,s}_{\kappa k_z}&=0.
\end{align}
\end{subequations}
For helicity,
\begin{subequations}\label{eq:QGTh}
\begin{align}
g^{h}_{\kappa\kappa}&=\frac{m^2\kappa^2}{4E^4p^2}+\frac{k_z^2}{4p^4},\\
g^{h}_{k_zk_z}&=\frac{m^2k_z^2}{4E^4p^2}+\frac{\kappa^2}{4p^4},\\
g^{h}_{\kappa k_z}&=\frac{m^2\kappa k_z}{4E^4p^2}-\frac{\kappa k_z}{4p^4},
&\Omega^{h,s}_{\kappa k_z}&=0.
\end{align}
\end{subequations}

These expressions provide local geometric fingerprints of the three projector families. The $K$ reduction is the only one with nonvanishing Berry curvature on the meridian, and Eq.~\eqref{eq:QGTK} agrees with the Abelian projection of the parent connection in Eq.~\eqref{eq:Kcurv}. By contrast, the $K_m$ and helicity reductions are curvature-flat on this particular two-dimensional pullback, although their quantum metrics are generically distinct and nonzero. Most importantly, vanishing meridional Berry curvature is not a criterion for global triviality: the helicity eigenlines carry nonzero flux through momentum spheres, as established in Sec.~\ref{sec:global}. The meridional quantum geometric tensor therefore characterizes the local embedding of the symmetry-resolved lines, while their extendibility and topology require the global projector analysis developed below.

\subsection{Gauge-covariant non-Abelianity and the massless limit}

The exact meridional reduction does not imply a corresponding reduction of the full three-dimensional parent connection. Once azimuthal variations are restored, the curvature acquires components along distinct Pauli directions. For example,
\begin{equation}
[\cF_{\kappa k_z},\cF_{\phi k_z}]
=\frac{\ii m\kappa}{2E^6}
\left[
\left(m+\frac{\kappa^2}{E+m}\right)\sigma_z
-\frac{\kappa k_z}{E+m}\sigma_r
\right].
\label{eq:Fcomm}
\end{equation}
Because the curvature transforms by conjugation, \(\cF_{ij}\to G^\dagger\cF_{ij}G\), the nonvanishing of such commutators is independent of the positive-energy frame. A rotationally invariant summary is obtained from the momentum-space ``magnetic'' curvature matrices
\begin{equation}
\cB_i\equiv\frac12\epsilon_{ijk}\cF_{jk}
=-\frac{1}{2E^3}
\left[m\sigma_i+\frac{p_i(\p\cdot\boldsymbol\sigma)}{E+m}\right],
\label{eq:Bcurv}
\end{equation}
for which
\begin{equation}
\mathfrak N
\equiv
-\frac12\sum_{i<j}\Tr\!\left([\cB_i,\cB_j]^2\right)
=
\frac{m^2\left(m^2+2E^2\right)}{4E^{12}}.
\label{eq:nonabdiag}
\end{equation}
The scalar \(\mathfrak N\) is invariant under positive-energy frame rotations and ordinary rotations of momentum space. It is a local diagnostic of the noncommuting curvature algebra rather than a topological invariant. For every \(m>0\), \(\mathfrak N>0\) at finite momentum, so the curvature components are not simultaneously diagonalizable at any finite momentum.

The massless limit is qualitatively different.  In the canonical Dirac frame, the helicity involution satisfies
\begin{equation}
D_i\gamma_h=\frac{m}{E}\,\partial_i\gamma_h.
\label{eq:Dgammah}
\end{equation}
Thus the helicity splitting is covariantly constant under the parent connection for $m=0$ at every $p>0$. This massless statement is gauge covariant. Consistently, setting \(m=0\) in Eq.~\eqref{eq:Fcart} gives
\begin{equation}
\cF_{ij}\big|_{m=0}
=-\frac{\epsilon_{ijk}p_k}{2p^4}(\p\cdot\boldsymbol\sigma),
\label{eq:masslessF}
\end{equation}
so, at each nonzero momentum, all curvature components are proportional to the helicity axis and therefore commute. The curvature algebra is thus pointwise Abelian in the helicity basis, and \(\mathfrak N\) vanishes identically.  The massless consequence of Eq.~\eqref{eq:Dgammah} strengthens the curvature-level result to covariant preservation of the helicity splitting under parallel transport.  It does not, however, imply the existence of a globally smooth helicity eigenvector frame: the nonzero Chern class still obstructs such a global gauge choice. For the massive problem, by contrast, Eq.~\eqref{eq:nonabdiag} shows that the exact Abelianization on a fixed-azimuth meridian is a property of that pullback rather than a global reduction of the parent non-Abelian geometry.

\section{Exact azimuthal holonomy}
\label{sec:holonomy}

A finite gauge-invariant characterization of the parent connection beyond the meridional reduction is provided by the Wilson loop along an azimuthal circle,
\begin{equation}
C_\phi:\qquad
\phi:0\longrightarrow2\pi,
\qquad
(\kappa,k_z)=\text{const.},
\qquad
\kappa>0,
\label{eq:Cphi}
\end{equation}
with
\begin{equation}
W_\phi=\mathcal P\exp\left(\ii\oint_{C_\phi}\cA\right).
\label{eq:Wdef}
\end{equation}
Under a smooth periodic change of positive-energy frame, $W_\phi$ transforms by conjugation; its spectrum and trace are therefore gauge invariant. Along $C_\phi$,
\begin{equation}
\cA_\phi(\phi)=a_\phi\sigma_r(\phi)+b_\phi\sigma_z,
\qquad
a_\phi=\frac{\kappa k_z}{2E(E+m)},
\qquad
b_\phi=-\frac{\kappa^2}{2E(E+m)}.
\label{eq:Aphiab}
\end{equation}
Because the radial Pauli matrix rotates with $\phi$, connection matrices at distinct points on the loop are generically noncommuting, so the path ordering in Eq.~\eqref{eq:Wdef} is essential. The loop can nevertheless be evaluated exactly by passing to a co-rotating spin frame. Writing the partial Wilson line as
\begin{equation}
\frac{\dd W}{\dd\phi}=\ii\cA_\phi(\phi)W,
\qquad
W(0)=\one_2,
\end{equation}
and introducing
\begin{equation}
R(\phi)=e^{-\ii\phi\sigma_z/2},
\qquad
\sigma_r(\phi)=R(\phi)\sigma_xR^\dagger(\phi),
\qquad
W(\phi)=R(\phi)V(\phi),
\end{equation}
reduces the evolution to
\begin{equation}
\frac{\dd V}{\dd\phi}
=\ii\left[a_\phi\sigma_x+\left(b_\phi+\frac12\right)\sigma_z\right]V.
\label{eq:Veq}
\end{equation}
The co-rotating frame is antiperiodic, $R(2\pi)=-\one_2$, and this endpoint factor must be retained because $R$ is not a periodic gauge transformation on $C_\phi$. Hence
\begin{equation}
W_\phi=-\exp\left\{2\pi\ii\left[a_\phi\sigma_x+\left(b_\phi+\frac12\right)\sigma_z\right]\right\}.
\label{eq:Wexactgenerator}
\end{equation}
The norm of the constant generator is
\begin{equation}
a_\phi^2+\left(b_\phi+\frac12\right)^2
=\frac{m^2+k_z^2}{4E^2}.
\label{eq:generatornorm}
\end{equation}
Defining
\begin{equation}
\mu=\sqrt{m^2+k_z^2},
\qquad
\Theta_\phi=\pi\left(1-\frac{\mu}{E}\right),
\label{eq:Theta}
\end{equation}
and choosing the eigenphases continuously so that they vanish when the circle collapses at $\kappa\to0$, one obtains the exact gauge-invariant spectrum
\begin{equation}
\operatorname{spec}W_\phi=
\left\{e^{+\ii\Theta_\phi},e^{-\ii\Theta_\phi}\right\}
\label{eq:Wspectrum}
\end{equation}
Equivalently,
\begin{equation}
\frac12\Tr W_\phi=\cos\Theta_\phi
=-\cos\left(\pi\frac{\mu}{E}\right),
\qquad
\det W_\phi=1.
\label{eq:Wtrace}
\end{equation}

The result has the expected contractible-loop limit: $\Theta_\phi\to0$ as $\kappa\to0$, while for a small circle
\begin{equation}
\Theta_\phi=\frac{\pi\kappa^2}{2\mu^2}+O(\kappa^4),
\end{equation}
consistent with the infinitesimal curvature flux. The Wilson loop is therefore geometric rather than topological; for $m>0$ the azimuthal circle may be contracted continuously to the momentum axis.

A minimal dynamical interpretation follows by considering the free Dirac Hamiltonian \(H(\p(t))\) with a slowly varying momentum parameter that traverses \(C_\phi\). The drive is assumed to be adiabatic relative to the particle--antiparticle energy separation, so that transitions out of the positive-energy doublet are suppressed. After removal of the common dynamical phase, the internal state follows the geometric transport \(W_\phi\) in the adiabatic limit.  At the initial and final point of the closed loop the \(K\) projectors coincide.  Since the constant generator in Eq.~\eqref{eq:Wexactgenerator} lies in the \((\sigma_x,\sigma_z)\) plane whereas \(\gamma_K(0)=-\sigma_y\), an initial \(K\) eigenstate has the following exact return and conversion probabilities for this geometric transport
\begin{equation}
P_{\eta\to\eta}^{(K)}
=\Tr\!\left[\pi_{K,\eta}(0)W_\phi\pi_{K,\eta}(0)W_\phi^\dagger\right]
=\cos^2\Theta_\phi,
\qquad
P_{\eta\to-\eta}^{(K)}
=\sin^2\Theta_\phi.
\label{eq:Kbranchconversion}
\end{equation}
Thus the Wilson angle controls a gauge-invariant branch-conversion probability for positive-energy geometric transport around the azimuthal loop.  This contrasts with fixed-azimuth meridional transport, for which Eq.~\eqref{eq:branchcovariant} is satisfied and the \(K\) branches are preserved.  Together with the noncommuting curvature algebra of Sec.~\ref{sec:geometry}, the finite conversion in Eq.~\eqref{eq:Kbranchconversion} shows that the exact meridional Abelianization does not extend to generic transport in the full parent bundle.

\section{Global extension properties of the conserved resolutions}
\label{sec:global}

Section~\ref{sec:doublet} established that the $K$, $K_m$, and helicity splittings are globally $SU(2)$-equivalent on their common regular domain $M=\{\kappa>0\}$. Throughout this section, the objects being extended are the internal momentum-space projectors $\Pi_{A,s}(\p)$ entering the Fourier synthesis~\eqref{eq:BesselFourier}, not fixed-angular-momentum Bessel basis vectors regarded as spatial states. The remaining question is an extension problem: what happens when the excluded momentum-space loci are restored? On a fixed nonzero momentum sphere $S_p^2$, the common domain is the punctured sphere $S_p^2\setminus\{N,S\}$. A conserved splitting may fail to extend to the missing points, it may extend smoothly and define a topologically trivial line bundle, or it may extend smoothly on the sphere while remaining obstructed from extension through the enclosed ball. The $K$, $K_m$, and helicity resolutions realize precisely these three possibilities, summarized in Table~\ref{tab:global}.

\begin{table}[tbp]
\centering
\small
\begin{tabularx}{\textwidth}{@{}l l l X@{}}
\toprule
reduction & splitting defect & behavior on $S_p^2$ & global outcome \\
\midrule
$K$ & $\kappa=0$ & fails at the two poles & no continuous extension of the prescribed projectors; no intrinsic full-sphere Chern number \\
$K_m$ & none for $m>0$ & smooth everywhere & smooth trivial; $C_{m,s}=0$ and the eigenlines extend through the enclosed momentum ball \\
helicity & $\p=0$ & smooth for every $p>0$ & smooth topological; $C_{h,s}=-s$ and the eigenlines cannot be extended through the origin \\
\bottomrule
\end{tabularx}
\caption{Global classification of the three conserved reductions of the positive-energy Dirac doublet on a fixed nonzero momentum sphere $S_p^2$. For $K$, the spectral defect lies on the base itself; for helicity, the only defect lies at the enclosed origin; and for $K_m$ no defect remains when $m>0$. The Chern-number sign follows the convention $\cA=\ii U^\dagger\dd U$ and the standard orientation of $S_p^2$.}
\label{tab:global}
\end{table}

The classification distinguishes the existence of a continuous extension from the topology of an extension when it exists. For $K$, the missing polar limits prevent completion of the punctured-sphere projector family. For $K_m$, the nonzero mass keeps the observable spectrum separated throughout the enclosed ball, permitting a smooth extension whose boundary line bundle is Chern trivial. For helicity, the projectors extend smoothly over the full nonzero momentum sphere, but their degree-one texture gives a nonzero first Chern class and obstructs continuation through the enclosed origin. These outcomes characterize distinct global extension properties of projector families that are already smoothly equivalent on their common regular domain.

\subsection{The \texorpdfstring{$K$}{K} reduction: nonextendability on the momentum axis}
\label{subsec:Kglobal}

For $\kappa>0$, the transverse-helicity operator has eigenvalues $\pm\kappa$ and defines two rank-one spectral projectors $\Pi_{K,\eta}$.  On the momentum axis, however,
\begin{equation}
K=0,
\end{equation}
so the two eigenvalues coalesce and the observable no longer selects a distinguished one-dimensional eigenspace.  Accordingly, the normalized operator
\begin{equation}
\Gamma_K=K/\kappa
\end{equation}
is undefined at $\kappa=0$.  The degeneracy removes the unique branch selection by $K$, but by itself does not decide whether the off-axis projectors admit a continuous limit.  That extension question is settled by their momentum dependence.

The obstruction can be seen directly on a fixed momentum sphere $S_p^2$.  In the canonical positive-energy frame,
\begin{equation}
\boldsymbol n_K(\phi)=(\sin\phi,-\cos\phi,0),
\label{eq:nK}
\end{equation}
so the limiting projector near either pole depends on the azimuthal direction of approach.  Hence $\Pi_{K,\eta}$ has no continuous extension to the north or south pole.  The $K$ branches therefore define line bundles only over the punctured sphere $S_p^2\setminus\{N,S\}$, not over the full $S_p^2$, and no intrinsic full-sphere first Chern number can be assigned to them.

Equation~\eqref{eq:nK} also shows that, in this particular frame, the internal axis traces the equator once as $\phi$ advances by $2\pi$.  This azimuthal winding serves only as a visualization of the transverse defect: the complex line bundle over the punctured sphere is topologically trivial, and the equatorial representation is not invariant under unrestricted smooth changes of positive-energy frame.  The gauge-invariant statement relevant to the present classification is the nonextendability of the spectral projectors themselves, established by their azimuth-dependent polar limits.  The spectral degeneracy at $\kappa=0$ is the point at which $K$ ceases to select the branches uniquely, but it is not by itself the proof of nonextendability.

\subsection{The transverse-separation reduction \texorpdfstring{$K_m$}{Km}: smooth extension and Chern triviality}

For the transverse-separation operator $K_m$,
\begin{equation}
K_m^2=m^2+\kappa^2\equiv\Lambda^2,
\end{equation}
so its two eigenvalues are $\pm\Lambda$ and remain separated by the observable spectral interval $2\Lambda\geq 2m$ for $m>0$ (an eigenvalue interval of $K_m$, not an energy splitting). The normalized operator
\begin{equation}
\Gamma_m=\frac{K_m}{\sqrt{m^2+\kappa^2}}
\label{eq:Gammamglobal}
\end{equation}
is consequently smooth throughout momentum space, including the axis and the origin, with $\Gamma_m|_{\kappa=0}=-\Sigma_z$. Since the positive-energy projector $P_+$ is also smooth for $m>0$, the rank-one projectors $\Pi_{m,s}=P_+(\one+s\Gamma_m)/2$ extend smoothly over every closed momentum ball $B^3_{p_0}$. The corresponding eigenlines therefore define complex line bundles over a contractible base. Their restrictions to the boundary sphere $S^2_{p_0}=\partial B^3_{p_0}$ are necessarily topologically trivial, and hence
\begin{equation}
C_{m,s}=0.
\label{eq:Cm}
\end{equation}
Thus smooth extendibility through the enclosed ball is the primary global statement; Chern triviality on the boundary sphere follows as a direct consequence.

The same conclusion is visible explicitly in the reduced Bloch texture. On $p=p_0$, with $\kappa=p\sin\theta$ and $k_z=p\cos\theta$,
\begin{equation}
\boldsymbol n_m
=\bigl(n_m^r\cos\phi,n_m^r\sin\phi,n_m^z\bigr),
\end{equation}
where $n_m^r,n_m^z$ are given in Eq.~\eqref{eq:gm}. For $m>0$,
\begin{equation}
n_m^z=-\frac{m(E+m)+\kappa^2}{\Lambda(E+m)}<0
\label{eq:nmznegative}
\end{equation}
throughout the sphere. The image of $S_p^2$ therefore lies entirely in the southern open hemisphere of the Bloch sphere and is explicitly null homotopic, in agreement with the ball-extension argument above. In the massless limit the lower bound on this observable spectral separation is lost: $\Lambda\to\kappa$, the two $K_m$ eigenvalues coalesce on the momentum axis, and $\Gamma_m$ ceases to be defined there. The role of the mass is therefore spectral and precise---it prevents the transverse splitting from closing and thereby permits the global extension of the $K_m$ eigenprojectors.

\subsection{Helicity: smooth spherical splitting and monopole topology}

Helicity realizes the third, and qualitatively different, global possibility.  The normalized operator~\eqref{eq:Gammah} is well defined for every $\p\neq0$ and has eigenvalues $\pm1$ on each fixed nonzero momentum sphere $S_p^2$. At $\p=0$ the normalized helicity operator is undefined; equivalently, the unnormalized operator $\boldsymbol\Sigma\cdot\p$ has eigenvalues $\pm p$ that become degenerate at the origin.  In the global canonical frame of the parent positive-energy bundle, the corresponding reduced projector is
\begin{equation}
\pi_{h,s}=\frac12\left(\one_2+s\widehat{\p}\cdot\boldsymbol\sigma\right),
\qquad
\boldsymbol n_h=\widehat{\p}
=(\sin\theta\cos\phi,\sin\theta\sin\phi,\cos\theta).
\label{eq:helicitysphereprojector}
\end{equation}
Thus the map $\boldsymbol n_h:S_p^2\to S^2_{\rm Bloch}$ has degree one.  With the Berry-connection convention adopted in Sec.~\ref{sec:geometry}, the branch curvature is
\begin{equation}
\Omega^{h,s}_{\theta\phi}=-\frac{s}{2}\sin\theta,
\end{equation}
and therefore
\begin{equation}
C_{h,s}
=\frac{1}{2\pi}\int_{S_p^2}\Omega^{h,s}
=-s.
\label{eq:Ch}
\end{equation}
The two helicity line bundles consequently carry opposite unit first Chern numbers, whose sum vanishes in the direct sum, consistently with the triviality of the parent rank-two bundle.  This is a momentum-space classification of helicity-selected eigenlines on a fixed sphere in a chosen inertial frame.  For $m>0$ it should not be confused with a decomposition of the massive one-particle Poincar\'e representation into Lorentz-invariant helicity sectors; massive helicity is not preserved by generic boosts~\cite{Polyzou2013}.

The nonzero Chern number gives a direct obstruction to extension through the enclosed ball.  If either helicity eigenline extended smoothly from $S_p^2$ to the enclosed ball $B_p^3$, its restriction to the boundary would have vanishing first Chern class because $B_p^3$ is contractible.  Equation~\eqref{eq:Ch} excludes such an extension.  The obstruction is therefore not a singularity on the sphere: the helicity projectors are smooth everywhere on $S_p^2$.  Rather, it is generated by the spectral defect at $\p=0$ enclosed by the sphere.  This differs from the transverse-separation reduction $K_m$, whose resolving operator remains nondegenerate throughout the same ball and whose boundary eigenlines are consequently Chern trivial.  With the spin-$\tfrac12$ normalization $h=s/2$, the massless limit of Eq.~\eqref{eq:Ch} agrees with the relation $C=-2h$ for massless helicity bundles~\cite{PalmerducaQin2025}.

\subsection{Common-domain equivalence and distinct global extension properties}

The explicit maps~\eqref{eq:gBK}--\eqref{eq:ghm} show that the three ordered splittings are globally unitarily equivalent on $M$. Restricting to a fixed nonzero momentum sphere therefore gives
\begin{equation}
\cL_{K,s}\simeq\cL_{m,s}\simeq\cL_{h,s}
\qquad
\text{over }S_p^2\setminus\{N,S\}.
\label{eq:puncturedequiv}
\end{equation}
The extension problem asks whether these equivalences can be continued across the excluded loci together with the corresponding projector families.

The $K$ case fails already at the level of existence: its eigenprojectors have azimuth-dependent limits at the two poles and therefore admit no continuous extension to the full sphere. The $K_m$ and helicity splittings both extend over $S_p^2$, so their completed line bundles can be compared directly. Suppose that the common-domain map $g_{h\leftarrow m}$ admitted a smooth extension $\widetilde g:S_p^2\to SU(2)$ satisfying
\begin{equation}
\pi_{h,t}=\widetilde g\,\pi_{m,s}\,\widetilde g^\dagger
\qquad\text{on }S_p^2.
\label{eq:globalUassume}
\end{equation}
Then $\widetilde g$ would induce an isomorphism $\cL_{m,s}\simeq\cL_{h,t}$, so their first Chern classes would have to coincide. The explicit values
\begin{equation}
C_{m,s}=0,
\qquad
C_{h,t}=-t=\pm1
\label{eq:ChernMismatch}
\end{equation}
exclude such an extension for every choice of $s$ and $t$. Thus the global $SU(2)$ equivalence on the common regular domain cannot be promoted to an equivalence of the completed $K_m$ and helicity eigenbundles.

The projective coordinate $\lambda$ does not affect this conclusion. It is only a local affine coordinate on the ray fiber $\mathbb{CP}^1$, so a divergence of $\lambda_{A,s}$ may be removed by passing to the reciprocal chart without changing the physical projector. Such a chart transition can neither create a missing polar limit nor alter a first Chern class. The hierarchy established by the three reductions is therefore
\begin{equation}
	\begin{gathered}
		\text{global } SU(2)\text{ equivalence on the common regular domain}\\
		\nRightarrow\quad \text{identical global extension properties}
	\end{gathered}
	\label{eq:localglobal}
\end{equation}
For $K$ the obstruction is nonextendability of the projector family itself; for $K_m$ and helicity it is the topological inequivalence of two smooth completions. This is the precise global content of the classification in Table~\ref{tab:global}.

\section{Discussion and conclusions}
\label{sec:discussion}

We have compared three conserved resolutions---transverse helicity $K$, transverse separation $K_m$, and helicity---inside the same positive-energy cylindrical Dirac doublet. Their ordered splittings are connected by explicit smooth, single-valued $SU(2)$ maps on the common regular domain $M=\{\kappa>0\}$, so they are equivalent as ordered bundle decompositions wherever all three are defined. Their extension properties differ when the excluded loci are restored: the $K$ projectors have azimuth-dependent polar limits and do not extend to the axis; for $m>0$ the $K_m$ projectors extend through the enclosed momentum ball and define Chern-trivial boundary lines; and helicity extends smoothly over every nonzero momentum sphere but carries $C_{h,s}=-s$, obstructing extension through the enclosed origin. The result is therefore a classification of the distinct global extension properties of three physically motivated cylindrical resolutions, rather than a classification of the complete free solution space.

The parent massive-Dirac bundle supplies a complementary geometric structure. On every fixed-azimuth meridian its connection and curvature lie in the $U(1)$ subalgebra generated by the projected $K$ axis, yielding exact meridional Abelianization and covariant preservation of the $K$ eigenlines. In the full three-dimensional momentum space the curvature algebra remains noncommutative for $m>0$, while the azimuthal Wilson loop is exactly solvable. For an adiabatic momentum cycle of the free Dirac Hamiltonian $H(\p(t))$, its Wilson angle gives the branch-resolved conversion probability in Eq.~\eqref{eq:Kbranchconversion}. These finite-transport results, together with the covariant projector formula~\eqref{eq:covariantbranchcurv}, distinguish the geometry of the unresolved rank-two doublet from that of its observable-selected eigenlines.

The role of cylindrical physics is to single out the three conserved resolutions that occur in concrete mode constructions: transverse-helicity modes in rotating fermionic systems, the transverse-separation sectors of boost-invariant Dirac-field quantization, and helicity-resolved Dirac--Bessel states in twisted-particle scattering~\cite{JiangLiao2016,Mihaila2006,Mihaila2009PRD,Serbo2015TwistedScattering,Bliokh2017Review,Karlovets2017JHEP}. The underlying non-Abelian massive-Dirac connection, the canonical-spin/helicity topology contrast, and observable-resolved subbundle topology have established precedents~\cite{ShankarMathur1994,ChenPangPuWang2014,PuYamamoto2018,Prodan2009,PalmerducaQinPhoton2024,PalmerducaQin2025}. The contribution here is their exact synthesis for these three cylindrical resolutions, including their common-domain intertwiners, distinct global extension properties, meridional reduction, and finite holonomy. These statements classify the free bulk spinor sector; applying them to finite rotating domains, interfaces or spatially varying masses and potentials, or particle-production backgrounds requires a separate check that the relevant operators preserve the full Hamiltonian domain and that the accessible momentum-space base retains the structure assumed here.

\acknowledgments
Q.G. acknowledges financial support from the National Natural Science Foundation of China through Grant No.~11874083.

\bibliographystyle{JHEP}
\bibliography{JHEP_v43_references}

\end{document}